\documentclass[twocolumn,superscriptaddress,prl]{revtex4}
\usepackage{graphicx,bm}
\newcommand{\be}{\begin{equation}}
\newcommand{\ee}{\end{equation}}
\newcommand{\bea}{\begin{eqnarray}}
\newcommand{\eea}{\end{eqnarray}}

\newcommand{\lp}{\left(}
\newcommand{\rp}{\right)}

\renewcommand{\phi}{\varphi}
\renewcommand{\epsilon}{\varepsilon}

\begin{document}

\title{
Fractional Quantum Hall Effect in Suspended Graphene: Transport Coefficients and Electron Interaction Strength
}

\author{D. A. Abanin}
\affiliation{Princeton Center for Theoretical Science, Princeton University, Princeton, New Jersey 08544, USA}
\author{I. Skachko}
\author{X. Du}
\author{E. Y. Andrei}
\affiliation{Department of Physics and Astronomy, Rutgers University, Piscataway, NJ 08855}
\author{L. S. Levitov}
\affiliation{Department of Physics, Massachusetts Institute of Technology, Cambridge, MA 02139, USA}
%% \date{\today}
\begin{abstract}
{\bf
Strongly correlated electron liquids which occur in quantizing magnetic fields 
reveal a cornucopia of fascinating quantum phenomena such as fractionally charged quasiparticles, anyonic statistics, topological order, and many others.
Probing these effects in GaAs-based systems, where electron interactions are relatively weak, requires sub-kelvin temperatures and record-high electron mobilities, rendering some of the most interesting states too fragile and difficult to access. This prompted a quest for new high-mobility systems with stronger electron interactions. Recently, fractional-quantized Hall effect was observed in suspended graphene (SG), a free-standing monolayer of carbon, where it was found to persist up to $T=10\,{\rm K}$. The best results in those experiments were obtained on micron-size flakes, on which only two-terminal transport measurements could be performed. Here we pose and solve the problem of extracting transport coefficients of a fractional quantum Hall state from the two-terminal conductance. We develop a method,
based on the conformal invariance of two-dimensional magnetotransport, and illustrate its use by analyzing the measurements on SG.  
From the temperature dependence of longitudinal conductivity,
extracted from the measured two-terminal conductance, 
we estimate the energy gap of quasiparticle excitations in the fractional-quantized $\nu=1/3$ state. The gap is found to be significantly larger than in GaAs-based structures, signaling much stronger electron interactions in suspended graphene. Our approach provides a new tool for the studies of quantum transport in suspended graphene and other nanoscale systems.
}
\end{abstract}

\maketitle

\section{Introduction}
Fractional quantum Hall effect is a remarkable manifestation of electron interactions in two spatial dimensions~\cite{Tsui82,Laughlin83}. The continued interest in this phenomenon is due to the large variety of fractionally quantized states and their rich physical properties~\cite{Pinzuk95}. In particular, quasiparticles of such states can carry a fraction of electron charge~\cite{Laughlin83,Reznikov97,Glattli97}, and obey anyonic statistics, rather than the usual bosonic or fermionic statistics~\cite{Wilczek83,Goldman}. It was predicted that non-abelian excitations may exist in the so-called $\nu=5/2$ state~\cite{MooreRead92}. So far most of experimental studies of these effects have been conducted in high-mobility GaAs-based structures, where the electron interactions are relatively weak. This limits the temperatures at which the fractional Hall effect can be observed to $T\lesssim 1\,{\rm K}$. 
Many of the states of interest, in particular the $5/2$ state, were found to be extremely fragile, which is reflected in the small energy gaps of elementary excitations. As a result, such states are only found in certain ultra-high-mobility GaAs structures~\cite{Willett87,Dolev08}, and probing their properties remains a challenge. This stimulated search for new two-dimensional systems with stronger electron interactions, which would host a larger variety of strongly correlated electron states.

Recently, much interest was generated by the realization of a new two-dimensional electron system in 
graphene, a one atom thick layer of crystalline carbon \cite{Novoselov04,CastroNeto09}.
One of the most exciting phenomena discovered in this material is the anomalous integer-quantized Hall effect (QHE) \cite{Novoselov05,Zhang05}.
Due to the massless Dirac character of the carrier dispersion, the cyclotron energy in graphene can be orders of magnitude larger than in systems with massive charge carriers in similar magnetic fields, reaching a few thousand kelvin in 10 tesla. As a result, the QHE in graphene persists up to room temperature \cite{Novoselov07}. 
Another interesting feature is that, being a semimetal, graphene hosts a family of Landau levels with particle/hole symmetry, resulting in a particle/hole-symmetric arrangement of QHE
plateaus in the transverse conductance, $\sigma_{xy}=\nu e^2/h$ ($\nu=\pm 2,\pm 6,\pm 10...$). A period-four regularity in these filling factors 
reflects the four-fold spin and valley degeneracy of Landau levels.

\begin{figure}
\includegraphics[width=3.5in]{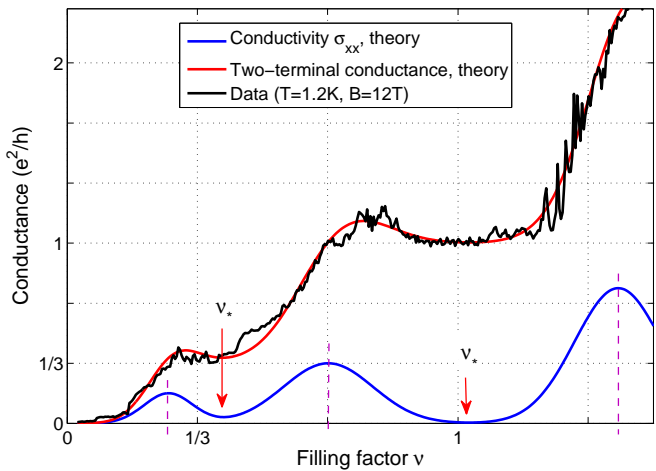}
\vspace{-1mm}
\caption[]{Theoretical fit (red) of the measured two-terminal conductance 
%% (red) shown together with the data from Ref.\cite{Du09} 
(black). The longitudinal conductivity $\sigma_{xx}$ (blue), used in the fit, has peaks at the plateau-to-plateau transitions and minima on the plateaus (vertical dashed lines and arrows, respectively). The best fit was obtained using the semicircle relation between $\sigma_{xy}$ and $\sigma_{xx}$, and treating the effective aspect ratio $L/W$, as well as the peaks' positions and widths, as fitting parameters as described in the text. Values at the minima, $\nu=\nu_*$, are used to estimate $\sigma_{xx}$ for incompressible QHE states.
}
\label{fig1}
\end{figure}

Graphene is a truly two-dimensional material, in which electron interactions are expected to be stronger than in other systems.
This should make interaction-induced QHE phenomena, such as the quantum Hall ferromagnetism (lifting of the valley and spin degeneracies of Landau levels) and the fractional quantum Hall effect (FQHE) easily observable \cite{Nomura06,Apalkov06,Toke06,Yang06}. 
Indeed, several integer QHE states outside the sequence $\nu=\pm 2, \pm 6...$ have been reported \cite{Zhang06,Jiang07,Checkelsky08}. These states, however, 
could only be observed
in very strong magnetic fields. The fragile character of these states was linked to substrate-related disorder which, by inducing spatial variation of electron density, suppresses the effects of interaction. 

A breakthrough, however, was achieved very recently in suspended graphene (SG) \cite{Bolotin08,Du08}. Electron mobility in SG, enhanced due to the absence of substrate, was found to approach the ballistic limit in micron-size flakes. Low disorder makes SG an ideal system for studying the effects of interaction. In recent measurements on suspended bilayer graphene, interaction-induced QHE states were observed in fields below 1 tesla \cite{Feldman09}. Most recently, fractional Hall effect was reported in single-layer SG, becoming visible in fields as low as 2 tesla \cite{Du09,Bolotin09}.

Best results in these experiments could be achieved with micron-size flakes, since in larger flakes the electron density becomes nonuniform due to SG sagging. However, in small flakes the standard four-probe measurement needed for separate measurement of the longitudinal and Hall conductivitiy is difficult to perform, because voltage probes in such small samples can short-circuit the Hall voltage. Observing the QHE using the standard Hall-bar geometry remains a challenge. In contrast, two-terminal measurements were found to consistently yield good results \cite{Du08,Feldman09,Du09,Bolotin09}. 
However, while pointing to the existence of QHE states, the results of a two-terminal measurement are not amenable to straightforward interpretation. Thus, no quantitative characteristics of the new QHE states were obtained in Refs.\cite{Du09,Bolotin09}.

\section{Deconvolving the two-terminal conductance}

In previous studies of QHE the longitudinal and Hall conductivities $\sigma_{xx}$ and $\sigma_{xy}$, obtained from Hall-bar measurements, have been 
the quantities of choice,
since their properties are most directly linked to the underlying physics. The width of the plateaus in $\sigma_{xy}$ and the depth of the corresponding minima in $\sigma_{xx}$ provide key information on localized states and on the electron interaction strength. Activation-like temperature dependence of $\sigma_{xx}$ can be used to extract the excitation gap, while deviations from simple activation behavior can reveal the nature of electron transport mechanism.

Since at present there are no reliable Hall-bar measurements in SG, it is tempting to use the two-terminal conductance for extracting the components of the conductivity tensor $\sigma_{xx}$ and $\sigma_{xy}$. However, the two-terminal conductance depends simultaneously on $\sigma_{xx}$, $\sigma_{xy}$ and sample geometry, and thus 'deconvolving' it requires an additional input. It was pointed out in Ref.\cite{Abanin08} that such input can be provided by the conformal invariance of the magnetotransport problem. In this approach, $\sigma_{xx}$ and $\sigma_{xy}$ are interpreted as a real and imaginary part of a complex number $\sigma=\sigma_{xx}+i\sigma_{xy}$, and thereupon the transport equations become conformally invariant. Applied to a rectangular two-lead geometry, 
theory yields a specific dependence of the two-terminal conductance on $\sigma_{xx}$, $\sigma_{xy}$ and the sample aspect ratio $L/W$. Interestingly, because of the conformal invariance, the same dependence describes the two-terminal conductance for an arbitrary sample shape, whereby the `effective aspect ratio' encodes the dependence on sample geometry.

Drawing on these observations about the role of conformal invariance, here we present a method for extracting transport coefficients $\sigma_{xx}$ and $\sigma_{xy}$ from the two-terminal measurements, and illustrate it using the data obtained as described in Ref.\cite{Du09}. 
%% The approach is 
%% applied to the 
%% conductance measured in 
%% data Ref.\cite{Du09}, giving a good description of the data in a fairly wide range of filling factors (see Fig.\ref{fig1}). 
The measurements were carried out on suspended graphene samples for temperatures ranging from $1\,{\rm K}$ to $80\,{\rm K}$ and fields up to 12 Tesla. The samples were fabricated from conventional devices mechanically exfoliated onto Si/SiO$_2$ substrates by removing the SiO$_2$ layer with chemical etching \cite{Du08}. In the final device the graphene sample is suspended from two Au/Ti pads which are split into two pairs one to apply the current and the other for probing voltage. The samples were typically 0.6 $\mu{\rm m} $ long and $1-3\,\mu{\rm m}$ wide and were probed in a regime where transport was ballistic.   

The 'half-integer' QHE in these samples is observed at fields as
low as 1T. The electron interaction effects become important at fields above 4T, resulting in new quantized plateaus which correspond to integer QHE
states at $\nu=0,\pm 1, \pm 3$, and a FQHE state at $\nu=1/3$ \cite{Du09}. The plateaus in conductance at these values of $\nu$ exhibit nearly perfect quantization expected for the two-terminal conductance in the QHE regime at fields above 8T. At moderate fields ($4-8$T) the plateaus are less well developed, which is likely to be a result of spatial density inhomogeneity. The data obtained at the highest field, $B=12$T, in which both the odd-integer and FQHE states are well developed, are best suited for our analysis.

As illustrated in Fig.\ref{fig1}, our approach provides a good description of the data in a fairly wide range of filling factors. Here we focus on the $\nu=1/3$ FQHE state and the $\nu=1$ interaction-induced QHE state, and extract $\sigma_{xx}$ on the plateaus. The temperature dependence of $\sigma_{xx}$, analyzed in terms of the activation transport mechanism, is used to estimate the excitation gap in these states. We find that the gap values which are two to five times greater than that measured in GaAs-based structures\cite{Tsui82,Boebinger85}.

Our approach, discussed in detail below, works best at temperatures which are neither too high, nor too low.
%% is subject to a number of limitations from both high and low temperatures.
The limitation from a high-temperature side stems from the nature of the semicircle relation, employed to describe the density dependence of $\sigma_{xy}$ and $\sigma_{xx}$ (see \cite{semicircle,Ruzin95} and references therein). This relation, which is also rooted in the complex-variable interpretation of magnetotransport, has been thoroughly tested in GaAs-based quantum Hall devices, both for integer and fractional QHE \cite{Hilke,Murzin}. It was found to work well deep in the QHE regime, i.e. at not too high temperatures, and less well at elevated temperatures. 

At the lowest temperatures, the conductance of our SG samples exhibits pronounced  mesoscopic fluctuations.
%% Another limitation, arising in small samples at the lowest temperatures, is due to the presence of mesoscopic fluctuations. 
These fluctuations dominate on the QHE plateaus,
% at low temperatures, 
rendering the description in terms of an average $\sigma_{xx}$ inadequate. Yet, as we show below, the approach based on the combination of conformal invariance and semicircle relation yields reasonably good results in a fairly wide range of intermediate temperatures, not too low and not too high.

At temperatures such that the mesoscopic fluctuations are not too prominent, we adopt an approximation in which the transport coefficients $\sigma_{xx}$ and $\sigma_{xy}$ take constant values throughout the sample. For a sample of a rectangular shape with ideal contacts at the opposite sides of the rectangle the solution to the transport problem has been long known~\cite{Lippmann56,Rendell81}. The result can be summarized in a compact form following Ref.~\cite{Lippmann56}, where it was derived with the help of conformal mapping. For a rectangle of length $L$ and width $W$ it is convenient to parameterize the aspect ratio as
\be
\ell=L/W=\frac{K(\sqrt{1-k^2})}{ 2K(k)}, 
\quad 0<k<1,
\ee
where $K$ denotes the complete elliptic integral of the first kind. The resistance of such a rectangle is then given by
\be\label{eq:resistanceRt}
R(\rho_{xx},\rho_{xy})=\sqrt{\rho_{xx}^2+{\rho}_{xy}^2} \frac{I(k',1)}{I(1,-1)}, 
\quad
k'=\frac{1}{k}>1
,
\ee
where the quantity $I(b,a)$ is defined as an integral
\bea\label{eq:elliptic}
&& I(b,a)=\!\int\limits_{a}^{b}\!\!\frac{d\xi}{|(\xi-1)(k'+\xi)|^{\alpha_-}
|(\xi+1)(k'-\xi)|^{\alpha_+}
},
\\
&& \alpha_\pm=1/2\pm \theta/\pi
,
\eea
where $\theta=\arctan(\rho_{xy}/\rho_{xx})$ is the Hall angle. Because of conformal invariance of the two-dimensional magnetotransport problem, Eq.(\ref{eq:resistanceRt}) also describes the two-terminal resistance for a sample of an arbitrary shape, with the dependence of $R$ on the sample geometry characterized by a single parameter $\ell$, the aspect ratio of an `equivalent rectangle' \cite{Abanin08}.

It is instructive to consider the behavior at large Hall angles, $\rho_{xx}\ll|\rho_{xy}|$, $\theta\to \pm\pi/2$, which is a regime relevant for our discussion below. 
Without loss of generality we can assume $\rho_{xy}>0$, in which case 
in the limit $\rho_{xx}\to 0$ we have $\alpha_-\to 0$ and $\alpha_+\to 1$. 
In this limit, the integrals in Eq.~(\ref{eq:resistanceRt}), dominated by the regions near $\xi=-1$ in $I(1,-1)$ and $\xi=k'$ in $I(k',1)$, diverge as $1/\alpha_-$. Extracting the leading contributions, we find that they are the same in both cases:
\be\label{eq:I11_2}
I(1,-1) \sim \frac1{(k'+1)\alpha_-},\quad I(k',1) \sim \frac1{(k'+1)\alpha_-} .
\ee 
Thus the ratio $I(k',1)/I(1,-1)$ tends to one, giving $R= |\rho_{xy}|$, which is the behavior expected in the dissipationless QHE state, when $\rho_{xx}=0$.
By carrying out the expansion in small $\alpha_-$ to next order, 
a contribution linear in $\rho_{xx}$ can be found (see Ref.~[\onlinecite{Lippmann56}]):
\be\label{eq:dR}
R=|\rho_{xy}| + \rho_{xx} g(\ell),\quad g(\ell)=\ln \frac{1-k}{2\sqrt{k}}.
\ee
The quantity $g(\ell)$ is positive for $\ell=L/W>1$, and negative for $\ell=L/W<1$; it vanishes at $\ell=1$. For $\ell\gg1$ (long and narrow sample) the function $g(\ell)$ is approximately linear: $g(\ell)\approx \ell$. For $\ell\ll 1$ (short and wide sample) the function $g(\ell)$ behaves as $1/\ell$. This behavior is consistent with the results expected for uniform current flow, $R(\ell\gg1)\sim\rho_{xx}\ell$, $R(\ell\ll1)\sim 1/(\sigma_{xx}\ell)$. Deviation from a quantized conductance value, described by Eq.(\ref{eq:dR}), offers a way to extract $\sigma_{xx}$ from two-terminal measurements.

Given the values of $\sigma_{xx}$, $\sigma_{xy}$ and $\ell$, the resistance $R$ can be obtained by numerical evaluation of the quantities in Eq.(\ref{eq:resistanceRt}). However, while the integral (\ref{eq:elliptic}) converges for all $-\pi/2<\theta<\pi/2$, the convergence is slow for large Hall angles, $\sigma_{xx}\ll|\sigma_{xy}|$, $\theta\to \pm\pi/2$, because of the power-law singularities of the integrand. Because of that, we found it more convenient to evaluate $R$ using another method which was developed by Rendell and Girvin~\cite{Rendell81}. In this approach, the current density is found as an exponential of certain infinite sums. The total current $I$ and source-drain voltage $V_{SD}$ are obtained by integrating the current density and electric field over appropriate contours, after which the resistance $R$ is found as the ratio, $R=V_{SD}/I$. The results obtained by this method are identical to those found from Eq.(\ref{eq:resistanceRt}). However, since the infinite sums giving current density converge rapidly, the numerics turns out to be substantially simpler than when Eq.(\ref{eq:resistanceRt}) is used directly. In what follows, we will use the approach of Ref.~[\onlinecite{Rendell81}] to evaluate the resistance.

There are several ways to use this approach for determining $\sigma_{xx}$. One is to focus on the plateaus, where $\sigma_{xx}$ is small and $\sigma_{xy}$ is quantized, $\sigma_{xy} = \nu e^2/h$. Expanding $G$ in the small ratio $\sigma_{xx}/|\sigma_{xy}|\ll 1$, the deviation from a quantized value can be expressed as $G = |\nu|e^2/h -g(\ell) \sigma_{xx}+O(\sigma_{xx}^2)$, where the coefficient $g(\ell)$ is a function of the aspect ratio only [see Eq.(\ref{eq:dR})]. 
Despite a conceptual simplicity of this approach, we found it difficult to implement, since the effective $L/W$ value may significantly deviate from the geometric aspect ratio of the sample, and thus should be treated as a fitting parameter \cite{Williams09}. 
Further, since $\sigma_{xx}$ and $\sigma_{xy}$   change with $\nu$, the conductance plateaus exhibit N-shaped distortions, rendering the deviation in $G$ from a quantized value unsuitable for accurate estimation of $\sigma_{xx}$.

\begin{figure}
\includegraphics[width=3.6in]{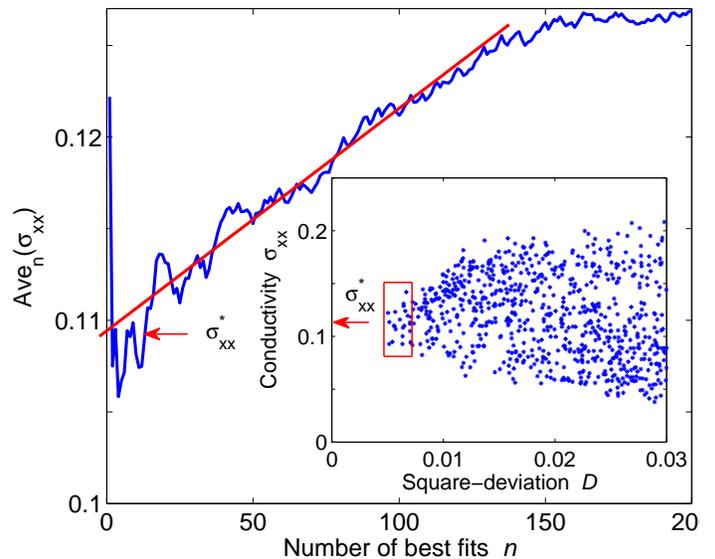}
\vspace{-1mm}
\caption[]{The statistical procedure used for extracting $\sigma_{xx}$ is illustrated for $\nu=1/3$ plateau of $T=6\,{\rm K}$ conductance trace (see Fig.\ref{fig2}a). Scanning a range of fitting parameters, including the aspect ratio $L/W$ and the widths and positions of gaussian peaks, Eq.(\ref{eq:gaussian}), yields several thousands fits. Quality of the fits is characterized by the standard deviation, Eq.(\ref{eq:standard_deviation}). The value at the  minimum, $\nu=\nu_*$, (see Fig.(\ref{fig1}) gives conductivity $\sigma_{xx}$ for each fit. The resulting conductivity was estimated by averaging the values of  $\sigma_{xx}$ over the group of fits within $30\%$ of the best fit, as shown by a box in the inset. Error bar was determined from the standard deviation of $\sigma_{xx}$ within this group of fits. Similar results can be found from the quantity ${\rm Ave}_n(\sigma_{xx})$, obtained by averaging  $\sigma_{xx}$ over $n$ best fits, extrapolated to $n=1$. 
}
\label{fig3}
\end{figure}

Considerably more reliable results can be obtained by focusing on the N-shaped distortions of QHE plateaus, since matching the theoretical model to an entire curve $G(\nu)$ puts a substantially more stringent constraint on the fitting parameters. The N-shaped features can be described by the density-dependent $\sigma_{xy}(\nu)$ and $\sigma_{xx}(\nu)$, which obey the semicircle relation \cite{Abanin08,Williams09}. 
For a plateau-to-plateau transition between incompressible filling factors $\nu_1<\nu_2$ this relation gives $\sigma_{xx}^2=(\sigma_{xy}-\nu_1)(\nu_2-\sigma_{xy})$ (in units of  $e^2/h$).  For fitting the conductance data shown in Fig.\ref{fig1}, which exhibits incompressible states at $\nu=0,\, 1/3,\, 1,\, 2$, we model the contribution of each QHE transition by a Gaussian,
\be\label{eq:gaussian}
\sigma_{xx}(\nu)=\frac12(\nu_2-\nu_1)e^{-A(\nu-\nu_c)^2}
,\quad \nu_1<\nu_c<\nu_2,
\ee 
and find the corresponding $\sigma_{xy}$ from the semicircle relation. This gives a contribution to the longitudinal and Hall conductivity of each of the relevant Landau levels or sublevels. The net conductivity $\sigma_{xy}(\nu)$ and $\sigma_{xx}(\nu)$, found as a sum of such independent contributions (blue curve in Fig.\ref{fig1}), is then used to calculate the dependence $G(\nu)=1/R$, where $R$ is given by
%% the two-terminal resistance, 
Eq.(\ref{eq:resistanceRt}).

We treat the $\sigma_{xx}$ peak positions and widths, as well as the effective ratio $L/W$, as variational parameters. As illustrated in Fig.\ref{fig1}, the Gaussian model with individually varying peak widths and positions 
provides a rather good description of the data. The sum of Gaussian peaks gives the quantity $\sigma_{xx}(\nu)$. The approach based on treating $L/W$ as a variational parameter, in general different from the actual sample aspect ratio, works rather well in the integer QHE regime \cite{Williams09}. It was conjectured \cite{Williams09} that variations in the best-fit value of $L/W$ account for the sample-dependent specifics of the current flow pattern such as those due to imperfect contacts and/or contact doping.

\section{Statistical analysis of the fitting procedure}

In order to extract $\sigma_{xx}$ in the incompressible 
$\nu=1/3$ and $\nu=1$ states, we analyzed a set of fits which best follow the data near the corresponding plateaus (Fig.\ref{fig1}). In both cases we found an optimal effective aspect ratio $L/W\approx 0.59$ (such deviation from the geometric aspect ratio, which in the sample \cite{Du09} was close to 0.4, is consistent with the results of Ref.\cite{Williams09}). Best fits were found from the square-deviation in the conductance averaged over a range of densities on the plateau and around it,
\be\label{eq:standard_deviation}
D=\int \lp G_{\rm theory}(\nu)-G_{\rm exp}(\nu)\rp^2 d\nu
,
\ee
In the case of the 1/3 plateau, a small interval near $\nu\approx 0.4$, which 
is probably related to another incipient FQHE feature, was excluded from the integral in Eq.(\ref{eq:standard_deviation}). 
The values at the minima, $\nu=\nu_*$ (see Fig.\ref{fig1}), were taken as an estimate of the longitudinal conductivity $\sigma_{xx}$ of incompressible QHE states. 

\begin{figure}
\includegraphics[width=3.5in]{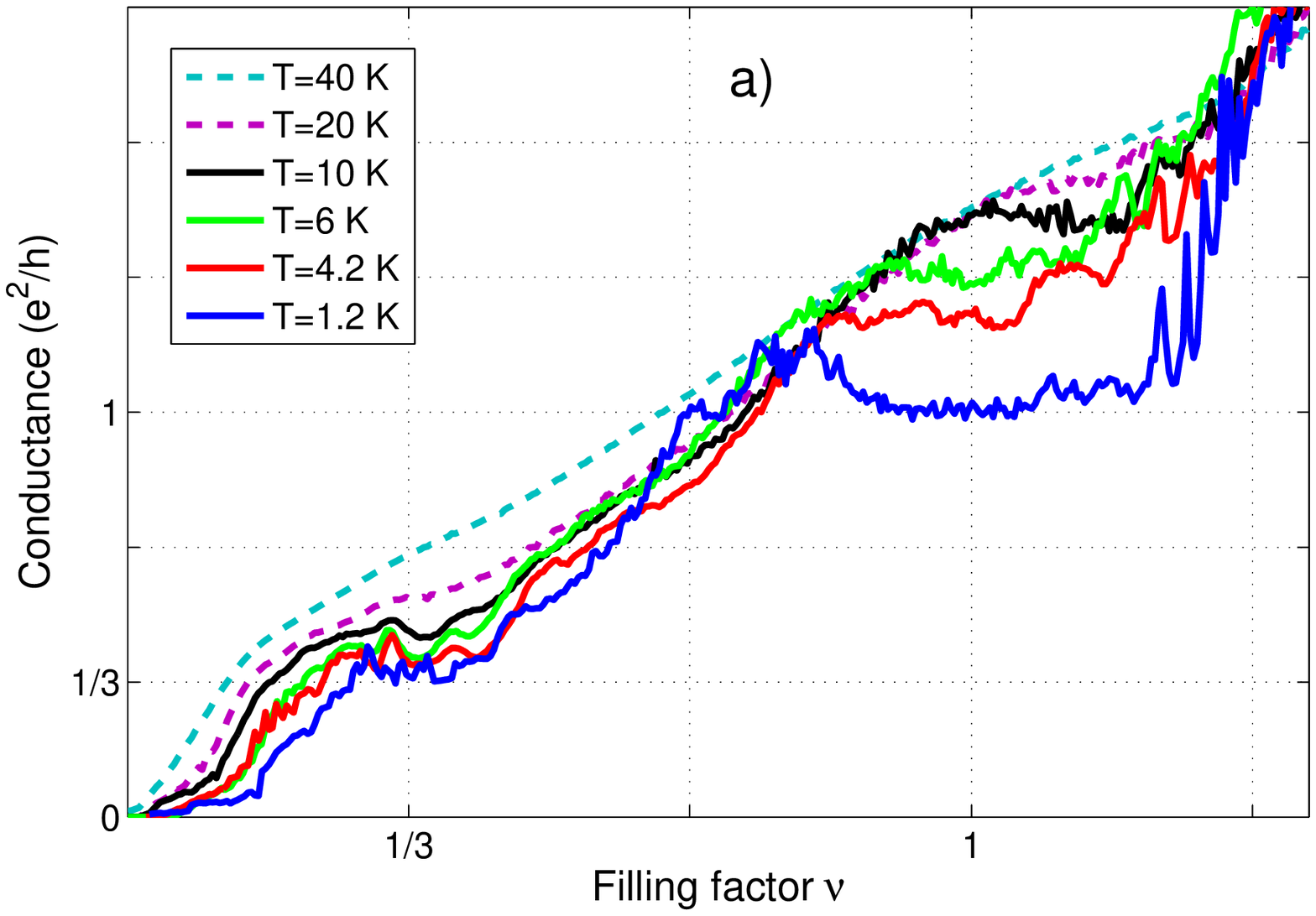}
\includegraphics[width=3.6in]{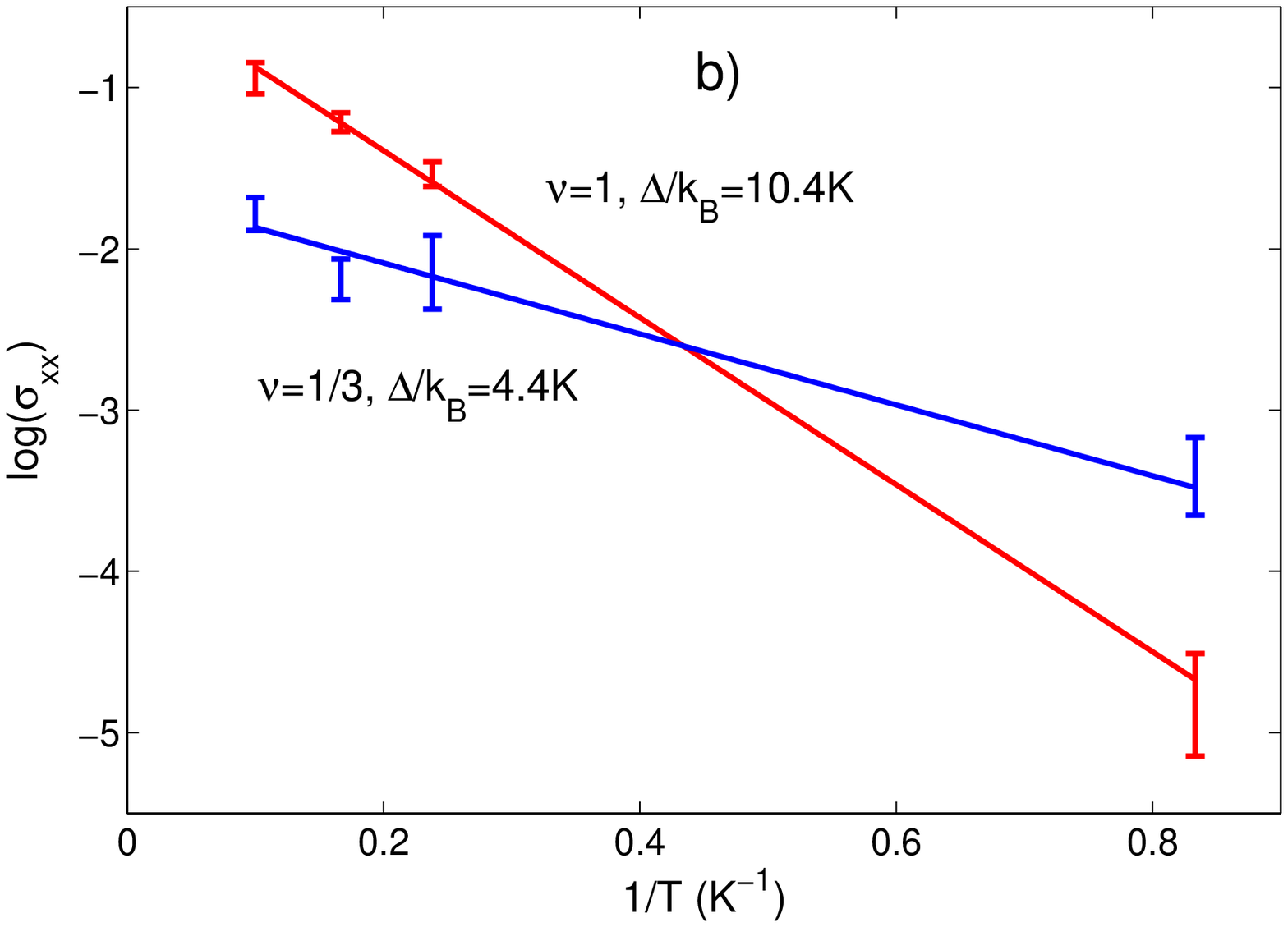}
\vspace{-1mm}
\caption[]{a) Temperature dependence of the two-terminal conductance 
%% (data from Ref.\cite{Du09})  used for extracting
from which $\sigma_{xx}$ on the $\nu = 1/3$ and $\nu= 1$ plateaus is extracted. 
b) Fits of the temperature dependence of $\sigma_{xx}$ to the activation model, $\sigma_{xx}\sim \exp(-\Delta/2k_BT)$. The best-fit values of the activation gap are $\Delta_{\nu=1}=10.4\,{\rm K}$ and $\Delta_{\nu=1/3}=4.4\,{\rm K}$. The procedure used to estimate error bars is outlined in Fig.\ref{fig3}.
}
\label{fig2}
\end{figure}

Search for the best fit was performed by optimizing fitting parameters, which include the positions and widths of Gaussian peaks, Eq.(\ref{eq:gaussian}), 
and the aspect ratio $L/W$. Our statistical analysis, based on comparing square-deviations $D$, Eq.(\ref{eq:standard_deviation}), for several thousands different, randomly chosen parameter values, is illustrated in Fig.\ref{fig3}. 
The value of $\sigma_{xx}$ was estimated by averaging over a group of fits within $30\%$ of the best fit, marked by a box in Fig.\ref{fig3} inset. 
Statistical error was estimated from the spread in the $\sigma_{xx}$ values found from the fits within this group.
The resulting error bar, displayed in Fig.\ref{fig2}b, is below $10\%$ at 10K, and increases to $20$-$25\%$ at 1.2K as a result of pronounced mesoscopic fluctuations developing at low temperatures. Alternatively, the value of $\sigma_{xx}$ could be determined be analyzing the mean value ${\rm Ave}_n(\sigma_{xx})$ taken over $n$ best fits, ordered according to their square-deviation $D$, as shown in Fig.\ref{fig3}. Linear extrapolation of ${\rm Ave}_n(\sigma_{xx})$ to $n=1$ yields results which are close to those found by averaging over the group of best fits, as described above, and has an advantage of being less subject to statistical error.

\section{Results and discussion} 

To extract $\sigma_{xx}$ at different temperatures, we used the 12 Tesla,
% data from Ref.\cite{Du09}, 
displayed in Fig.\ref{fig2}a. On the plateaus, the conductance increases with temperature faster than off the plateaus, consistent with the behavior expected in the QHE regime. Both plateau features remain visible up to $T\sim 20\,{\rm K}$. For the highest temperatures, however, the density-dependent conductivity $\sigma_{xx}(\nu)$ exhibits very shallow minima on the plateaus, giving rise to a large statistical error in fitting. We attribute this behavior to the temperatures $T\gtrsim 20\,{\rm K}$ lying outside the range of applicability of the semicircle model, similar to the highest temperatures in Refs.\cite{Hilke,Murzin}. Thus we exclude the $T= 20,\ 40\,{\rm K}$ traces from the analysis. Statistical error, estimated from fluctuations of $\sigma_{xx}$ over a group of best fits, as discussed above, gives the error bars shown in Fig.\ref{fig2}b. Larger statistical error found for $T=1.2\,{\rm K}$ reflects growth of mesoscopic fluctuations, which dominate the transport on the QHE plateaus at the lowest temperatures.

The conductivity values on the plateaus, obtained from the conductance traces with $T=1.2$, 4.2, 6 and 10 K, were analyzed in terms of the activation behavior, $\sigma_{xx}\sim \exp(-\Delta/2k_BT)$ (see Fig.\ref{fig2}b). The best fit values of the energy gap are $\Delta/k_B = 10.4\,{\rm K}$ for $\nu=1$ and $\Delta/k_B = 4.4\,{\rm K}$ for $\nu=1/3$. The relatively high energy scale 
is consistent with the disappearance of the corresponding plateaus at about 20K. We also considered fits to the variable-range-hopping dependence, $\sigma_{xx}\sim \exp\lp -(T_*/T)^\eta\rp$, $\eta=1/2$, but have not found a discernible statistical advantage over the activation dependence.

It is instructive to compare these results with the activation gap measured in the $\nu=1/3$ state in GaAs structures. In the first experiment, in which the FQHE at $\nu=1/3$ was discovered~\cite{Tsui82}, the temperature dependence of $\sigma_{xx}$ was weak, and the activation gap could not be determined.
% temperature dependence 
% %% of the order $e^2/h$ 
% down to the lowest temperature, $T=0.48\,{\rm K}$.
%% , likely indicating the activation gaps comparable to $T_0$, which is a factor of 10 smaller than the gap we extracted for SG.
Subsequent measurements on samples of higher
mobility~\cite{Boebinger85} revealed activated behavior of
conductivity at temperatures
 $T>T_*$, with $T_*$ of the order of $0.1-1\,{\rm K}$, depending on the field. 
The activated dependence crossed over to the variable-range-hopping behavior at lower temperatures, indicative of the localization of quasiparticles. 
Despite very high mobility of GaAs structures, the extracted gap value, which was about $2\,{\rm K}$ at $B=12\,{\rm T}$, is about 2.5 times lower than the value for SG obtained above, indicating a more robust nature of FQHE in graphene. 

Given that the FQHE features in SG remain clearly visible up to 10-20\,K (see Fig.\ref{fig2}a), whereas in the measurements on GaAs the FQHE persisted only up to about 2\,K, one may expect subsequent measurements on cleaner SG samples to revise the gap values. Another indication that our estimate of the gap is merely a lower bound arises from comparison to theoretically predicted values. Theoretical estimate for $\nu=1/3$ in SG gives $\Delta_{\nu=1/3}=\alpha e^2/\kappa \ell_B$, where $\alpha\approx 0.1$~\cite{Apalkov06}. taking the dielectric constant $\kappa=5.24$, which is the RPA result for intrinsic screening function of graphene~\cite{Gonzalez99}, for $B=12\,{\rm T}$ we obtain $\Delta_{\nu=1/3}=42\,{\rm K}$.

This situation can be contrasted with GaAs, where the theoretical value, $\Delta_{\nu=1/3}\approx \tilde\alpha e^2/\epsilon \ell_B$, with $\tilde\alpha\approx 0.03$ and $\epsilon=12.8$, is only a factor of 2.5 greater than experimental value measured at $B=12\,{\rm T}$~\cite{Boebinger85}. A relatively small value of the prefactor $\tilde\alpha$ accounts for the effect of finite width of GaAs quantum wells~\cite{DasSarma86}, which makes the short-range interactions weaker than in a truly two-dimensional system such as graphene. In higher fields, $B\sim 20\,{\rm T}$, the theoretical limit in GaAs has been nearly reached~\cite{Boebinger85}, due to the enhanced role of interactions compared to disorder.

We therefore believe that 
% a relatively low value 
the departure of the gap inferred in present work from theoretical predictions reflects the effect of disorder present in the system. In particular, it was pointed out that rippling of a suspended graphene sheet may result in formation of localized midgap states~\cite{CastroNeto09}. Yet, similar to the case of GaAs, the effect of disorder should become weaker at higher magnetic fields. This would bring the FQHE gaps closer to the very large theoretically predicted values, roughly a factor of 20 greater than those in GaAs. Realizing FQHE with larger gaps should also be possible at lower magnetic fields once the sample quality is improved.

\section{Summary}

In this work, 
%% In summary, 
we developed a general method for extracting transport coefficients in a QHE state from the two-terminal conductance. We demonstrate that, while the two-terminal conductance depends in a fairly complicated way on the sample geometry, as well as on $\sigma_{xx}$ and $\sigma_{xy}$, a reliable procedure for determining these values can be developed. This type of analysis is made possible by constraints on this quantity arising from conformal invariance of the two-dimensional magnetotransport and the semicircle relation between $\sigma_{xx}$ and $\sigma_{xy}$ in a QHE state. We apply our approach to analyze the fractional and integer QHE states in suspended graphene flakes, where, because of small sample size, only two-terminal measurements can be performed \cite{Du09,Bolotin09}. 

We estimate the energy gap of the quasiparticle excitations in the $\nu=1/3$ FQHE state by analyzing the temperature dependence of $\sigma_{xx}$. The gap is found to be significantly larger than in semiconducting systems, signaling stronger electron interactions in graphene. From a comparison to measurements in GaAs and to theoretical estimates, we conclude that the effects of electron interactions in current SG samples, despite being somewhat masked by disorder, are stronger than in high-quality GaAs structures. We expect that the future experiments on cleaner SG will reveal exceptionally robust FQHE, with gaps reaching a few tens of Kelvin. Given the richness and diversity of FQHE phenomena in GaAs structures, it is natural to expect that SG will be shown to host new types of FQHE, which are not observable in semiconducting structures because of the much weaker Coulomb interactions.

DA thanks Nordic Center for Theoretical Physics (NORDITA) for hospitality during program ``Quantum Hall Physics", where part of this work was completed. LL acknowledges ONR support under N00014-09-1-0724. EA acknowledges DOE support under DE-FG02-99ER45742  and partial NSF support under NSF-DMR-045673.

\end{document}